\documentclass{ametsocV6.1}

\usepackage{amsmath,amsfonts,amssymb,bm}
\usepackage{mathptmx}
\usepackage{newtxtext}
\usepackage{newtxmath}
\usepackage{lineno}
\usepackage{bm}
\linenumbers
\usepackage[hidelinks]{hyperref}

\usepackage{siunitx}

\DeclareSIUnit{\days}{days}

\title{Interpreting Observed Interactions between \\ Near-Inertial Waves and Mesoscale Eddies}

\authors{Scott Conn,\aff{a}\correspondingauthor{Scott Conn, sconn@caltech.edu}
Joseph Fitzgerald\thanks{Current affiliation: Memorial University of Newfoundland, St John's, Canada},\aff{a}
J\"orn Callies\aff{a}}

\affiliation{\aff{a}{California Institute of Technology, Pasadena, California}}

\abstract{The evolution of wind-generated near-inertial waves (NIWs) is known to be influenced by the mesoscale eddy field, yet it remains a challenge to disentangle the effects of this interaction in observations. Here, the model of Young and Ben Jelloul (YBJ), which describes NIW evolution in the presence of slowly evolving mesoscale eddies, is compared to observations from a mooring array in the Northeast Atlantic Ocean. The model captures the evolution of both the observed NIW amplitude and phase much more accurately than a slab mixed layer model. The YBJ model allows for the identification of specific physical processes that drive the observed evolution. It reveals that differences in the NIW amplitude across the mooring array are caused by the refractive concentration of NIWs into anticyclones. Advection and wave dispersion also make important contributions to the observed wave evolution. Stimulated generation, a process by which mesoscale kinetic energy acts as a source of NIW potential energy, is estimated to be \qty{20}{\micro\watt\per\meter\squared} in the region of the mooring array, which is two orders of magnitude smaller than the global average input to mesoscale kinetic energy and likely not an important contribution to the mesoscale kinetic energy budget in this region. Overall, the results show that the YBJ model is a quantitatively useful tool to interpret observations of NIWs.}

\begin{document}
\nolinenumbers

\maketitle

%
%
%

%








\section{Introduction}\label{intro}
Near-inertial waves (NIWs), internal waves with a frequency close to the inertial frequency~$f$, are resonantly excited by atmospheric winds exerting a stress on the ocean's surface. It has long been recognized that these waves can interact with mesoscale eddies and that this interaction may be important in the life cycle of wind-generated NIWs. Observational evidence of NIW--mesoscale interactions is accumulating, but interpreting the observed NIW evolution in the presence of mesoscale eddies remains challenging. Here, we employ the theoretical framework of \citet[][from hereon YBJ]{young1997} to identify and interpret such interactions in mooring observations in the Northeast Atlantic Ocean.

NIWs are associated with a large vertical shear, which can result in shear instabilities and vertical mixing \citep[for a review, see][]{alford2016}. These shear instabilities are a key mechanism by which atmospheric storms can cause the surface mixed layer to deepen. \citet{jochum2013} showed that an improved representation of NIWs in a climate model led to a deepening of the mixed layer on average, which in turn resulted in significant changes in sea surface temperatures, winds, and precipitation. 

The amount of NIW shear in the upper ocean, and as a consequence the amount of mixed layer deepening, depends on both the energy input into the NIW band and how rapidly NIWs propagate to depth. The vertical propagation originally presented a paradox. The vertical group velocity of NIWs varies as $\kappa^2$, where $\kappa$ is the horizontal wavenumber. The atmospheric storms that generate NIWs are typically $O$(\qty{1000}{\kilo\meter}) in size and generate NIWs with a similar scale. Estimates of the vertical group velocity based on this horizontal scale are much too slow to explain the observed decay of NIWs in the mixed layer and propagation to depth \citep{dasaro1995a}. A reduction in the horizontal scale of the waves is required to obtain a group velocity that matches observations. There are two possible mechanisms by which this is thought to occur: (i)~meridional variations in~$f$ cause an increase in the meridional wavenumber \citep[$\beta$-refraction;][]{gill1984}, or (ii)~interactions with the mesoscale circulation can imprint structure onto the wave field at the scale of mesoscale eddies \citep[$\zeta$-refraction; YBJ;][]{kunze1985}. The presence of the former alone is enough to cause propagation of NIWs out of the mixed layer, with the NIW kinetic energy in the mixed layer decaying as $t^{-3/2}$ \citep[][]{moehlis2001radiation}. The latter process causes concentration of NIW energy into anticyclones and subsequent downward propagation in inertial ``chimneys'' \citep{lee_inertial_1998} or ``drainpipes'' \citep{asselin2020penetration}.

There is a long history of NIW observations from \textit{in situ} measurements. The Ocean Storms Experiment \citep{dasaro1995a} was a groundbreaking observational campaign to study the life cycle of wind-generated NIWs. The experiment tracked the response of the surface ocean to strong wind forcing in a region with weak mesoscale eddies. It was found that $\beta$-refraction was the dominant process driving the observed evolution of the meridional wavenumber of the waves. The process of $\zeta$-refraction was found to have much less of an effect than expected \citep{dasaro1995c}, which YBJ argued was due to strong dispersion in a weak eddy field. More recently, \citet{thomas2020} used ship-based observations of NIWs in a strong dipole vortex to estimate the rate of change in time of the NIW horizontal wavenumbers. Here, the expected change in horizontal wavenumbers from $\zeta$-refraction was consistent with the data. The observations further showed a NIW beam at depth, indicating that the scale reduction had allowed the NIWs to propagate vertically out of the mixed layer. \citet{essink2022} studied typhoon-forced NIWs in the Kuroshio Current. They observed the trapping of NIWs in a region of anticyclonic vorticity, followed by downward propagation. They also measured the vertical structure of turbulence and showed how this was influenced by NIW dynamics. \citet{yu2022} investigated the interaction of NIWs and mesoscale eddies in observational data of NIWs from the Northeast Atlantic (the same data as we will use below). They showed that regions of elevated NIW kinetic energy are statistically associated with mesoscale anticyclones and that the submesoscale vorticity exerted little control on the horizontal concentration of NIWs and the subsequent propagation to depth. These studies, however, often focus on a single physical process (usually refraction) and many make assumptions about the NIW dynamics that may not be universally justified. Here, we propose that the YBJ model is a general theoretical framework that, when applied to observations, allows us to understand the multiple physical processes that govern NIW evolution.

It has also been proposed that NIWs not only react to the presence of mesoscale eddies but feed back on the eddies and affect their evolution. Approximately 80$\%$ of the ocean's kinetic energy exists as mesoscale motions \citep{ferrari2009}. The geostrophic constraint on mesoscale eddies traps energy at large scales, and it is not entirely clear how the energy input into mesoscale motion is balanced by dissipation \citep{muller_routes_2011}. A number of mechanisms by which mesoscale eddies lose energy are known, including dissipation in bottom boundary layers \citep[e.g.,][]{arbic_baroclinically_2004}, the generation of dissipative lee waves \citep[e.g.,][]{nikurashin_routes_2013}, energy loss near western boundaries \citep[e.g.,][]{zhai_significant_2010}, and the top drag arising from the current dependence of the wind stress \citep[e.g.,][]{dewar_effects_1987,renault2016}. The extraction of energy from mesoscale eddies by NIWs presents another possibility \citep{xie2015, rocha2018}. Given the great importance of mesoscale eddies to the transport of heat and carbon \citep[e.g.,][]{jayne_oceanic_2002,gnanadesikan_isopycnal_2015}, even small changes \citep[see discussion in][]{asselin2020penetration} to the mesoscale eddy field caused by NIWs may be significant to the impact of the ocean on climate. 

To understand the propagation of NIWs and by extension the role they play in upper-ocean mixing, we need to understand the dynamics governing their evolution. Slab mixed layer models \citep[][from hereon PM]{pollard1970} are a commonly used tool to model NIW evolution \citep[e.g.,][]{d1985energy,alford2001,alford2020,guan2014observed}. PM showed that a slab model can reproduce key features of the NIW evolution observed by moorings. With the horizontal NIW velocity denoted by $(u,v)^\mathrm{T}$, the PM model can be written as:
\begin{subequations}
    \begin{align}
    \partial_tu - fv = -ru + \frac{\tau^x}{\rho_w H_m},\label{eq:PM1}\\
    \partial_tv + fu = -rv + \frac{\tau^y}{\rho_w H_m},\label{eq:PM2}
    \end{align}
\end{subequations}
where $r^{-1}$ is a decay timescale, $\rho_w$ is the density of water, $H_m$ is the mixed layer depth, and $\bm{\tau}$ is the wind stress. This model does not explicitly represent any physical processes affecting the evolution of NIWs other than the wind forcing and Coriolis effect. All other processes are subsumed in the linear drag, which can be thought of as a parameterization primarily for the propagation of NIWs out of the mixed layer. Observations indicate that this process is slow relative to the inertial frequency \cite[e.g.,][]{dasaro1995a}, and so we require $r\ll f$. Other than this restriction, $r$ is a tunable parameter.

Despite its successes in capturing some aspects of the observed NIW signal, the PM model cannot explain the propagation of NIWs out of the mixed layer. If the NIW field is initially uniform, it will remain so. The model captures neither $\beta$-refraction nor the interaction with mesoscale eddies.

A more general framework to understand the evolution of NIWs was devised by YBJ by assuming a time scale separation between the fast waves and the slowly evolving mesoscale flow. In the YBJ framework, the horizontal NIW velocity is first complexified (written as $u+iv$). Since NIWs have a frequency close to $f$, it is convenient to write this complexified velocity as $u+iv = e^{-ift}\partial_zM(x,y,z,t)$. The function $\partial_zM$ describes the slow evolution of the envelope that modulates the NIW phase and amplitude. The YBJ equation describes how $\partial_zM$ evolves in the presence of prescribed geostrophic mesoscale eddies. On an $f$-plane, the equation reads
\begin{equation}
    \partial_{zzt}M + J(\psi,\partial_{zz}M)+\frac{iN^2}{2f}\nabla^2 M +  \frac{i\zeta}{2}\partial_{zz}M = \partial_{zz}\mathcal{F} - \nu\nabla^4\partial_{zz}M\label{eq:standard_YBJ},
\end{equation}
where $\psi$ is the geostrophic streamfunction, $\nabla^2=\partial_x^2+\partial_y^2$~is the horizontal Laplacian operator, $\zeta=\nabla^2\psi$ is the geostrophic vorticity, $N^2$~is the stratification, $\mathcal{F}$~is a forcing term that represents the momentum flux due to the surface wind stress, and $\nu$~is a hyper-diffusivity included for numerical stability (see Section~\ref{models}). In~\eqref{eq:standard_YBJ} and throughout this paper, we assume the geostrophic flow to be barotropic (i.e., $\psi$ is independent of depth), although baroclinicity in the geostrophic eddy field can also be taken into account. Unlike the PM model, the YBJ equation does not have a tuneable parameter. 

The second term in \eqref{eq:standard_YBJ} represents advection of NIWs by the mesoscale flow. The third term represents changes in the NIW field due to dispersion. The presence of the dispersion term means that an initially localized wave packet will spread out as time progresses. The fourth term is responsible for the process of $\zeta$-refraction. This term sets into motion the imprinting of mesoscale structure onto an initially horizontally uniform wave field. This $\zeta$-refraction shifts the phase of the NIWs, which we can see by neglecting all other terms in the YBJ equation:
\begin{equation}
    \partial_{zzt}M = -\frac{i\zeta}{2}\partial_{zz}M.
\end{equation}
Assuming a steady vorticity field yields solutions of the form $\partial_{zz}M = C(x,y,z)e^{-i\zeta t/2}$. Spatial heterogeneities in the mesoscale vorticity will result in spatial heterogeneities in the NIW phase. The dispersion term in the full YBJ equation~\eqref{eq:standard_YBJ} then acts on these phase gradients and fluxes energy into anticyclones and out of cyclones. This can be seen from the YBJ kinetic energy budget (Appendix~A):
\begin{equation}
    \partial_t\mathcal{K} + J(\psi,\mathcal{K}) + \nabla\cdot\mathbf{F}_\mathcal{K} + \partial_zG_\mathcal{K} = \gamma_\mathcal{K}^F + d_\mathcal{K}, \label{eq:KE_budget}
\end{equation}
where $\mathcal{K}$ is the NIW kinetic energy density, $\mathbf{F}_\mathcal{K}$ and $G_\mathcal{K}$ are the horizontal and vertical energy fluxes due to dispersion, $\gamma_\mathcal{K}^F$ is the energy input by the forcing, and $d_\mathcal{K}$ is the dissipation due to the hyperviscosity term. The horizontal energy flux $\mathbf{F}_\mathcal{K}$ can be expressed in terms of gradients of the phase~$\Theta$ of $M$ \citep{rocha2018}:
\begin{equation}
    \mathbf{F}_\mathcal{K} = \frac{N^2|M|^2}{2f}\nabla\Theta.
    \label{eq:fluxphase}
\end{equation}
The spatial heterogeneities that the mesoscale vorticity~$\zeta$ induces in the NIW phase thus cause a transfer of NIW energy in the horizontal. Once this $\zeta$-refraction has imprinted the horizontal structure of eddies onto an initially uniform NIW field, advection will also act on the resulting gradients and stir the NIW field. Dispersion remains important and helps waves escape straining regions \citep{rocha2018}. The scale reduction also accelerates the propagation of NIWs out of the mixed layer. This is represented in the YBJ energy budget~\eqref{eq:KE_budget} through the vertical flux~$G_\mathcal{K}$, which requires horizontal gradients in the NIW field to produce energy fluxes to depth (Appendix~A).

The energy input into the NIW band by the winds~$\gamma_\mathcal{K}^F$ is known as the NIW wind work. The difference between the NIW wind work and the NIW kinetic energy that propagates out of the surface layer is the energy that is available for NIW mixing in the surface ocean. Both processes therefore influence the mixed layer depth. There has been extensive effort to estimate the NIW wind work. It can be calculated directly from concurrent observations of winds and NIW surface currents, but this is possible in a few locations only. Slab models have been used to obtain global estimates of the NIW wind work \citep{alford2001} but these are suspected to be overestimates, primarily because these models poorly represent the various processes that cause NIWs to leave the mixed layer. This can include the vertical propagation discussed above, but the employed models often also do not represent dissipation of NIWs by conversion to turbulent kinetic energy, which would reduce the projection of the wind stress onto the waves \citep{plueddemann2006observations,alford2020}. High-resolution ocean models can also be used to estimate the wind work, but here problems can arise from limitations in the reanalysis data used to force the models. For example, \citet{flexas2019} showed that the NIW wind work was poorly represented in a high-resolution ocean model due to the wind forcing missing variability on scales less than 6~hours and 15~km. The power input is larger in high-resolution coupled simulations but still substantially lower than observational estimates from slab models \citep{von_storch_wind_2023}.

The mesoscale can act as a source of wave potential energy in a process known as stimulated generation. In unforced and inviscid YBJ dynamics, the wave kinetic energy is conserved but NIWs can gain or lose potential energy through interactions with the mesoscale. In the YBJ framework, the domain-integrated potential energy is not conserved (Appendix~B):
\begin{equation}
    \partial_t\langle\mathcal{P}\rangle = \Gamma_\mathcal{P}^R + \Gamma_\mathcal{P}^A + \mathcal{D}_\mathcal{P},\label{eq:domain_integrated_PE_budget}
\end{equation}
where $\mathcal{P}$ is the potential energy density, $\Gamma_\mathcal{P}^R$ is the production of potential energy by refraction, $\Gamma_\mathcal{P}^A$ is the production of potential energy by advection, and $\mathcal{D}_\mathcal{P}$ is the potential energy dissipation by hyperviscosity. The formation of horizontal structure in an initially uniform NIW causes an increase in NIW potential energy via $\Gamma_\mathcal{P}^R$ and $\Gamma_\mathcal{P}^A$. In the first phase of interaction, $\Gamma_\mathcal{P}^R$ dominates as the waves have the mesoscale structure imprinted on them. At later times, $\Gamma_\mathcal{P}^A$ becomes the most important term. An extension of YBJ describing the coupled evolution of NIWs and a quasi-geostrophic mesoscale field was derived by \citet{xie2015} \citep[see also][]{wagner2016}. This extension shows that the refractive and advective sources of potential energy to the NIWs appear as sinks in the mesoscale energy budget. This allows us to interpret $\Gamma_\mathcal{P}^R$ and $\Gamma_\mathcal{P}^A$ as energy transfers from the mesoscale, despite the fact that we impose the mesoscale field in the simulations described below. Theoretical and numerical studies \citep{xie2015,rocha2018,asselin2020penetration} have investigated the process of stimulated generation, yet its importance in the real ocean remains poorly constrained.

Several attempts have been made to reconcile available observations with our theoretical understanding of NIW--mesoscale interactions. Work on the NIW--mesoscale interaction prior to YBJ had been based on ray tracing theory, which additionally assumes that the waves have much shorter spatial scales than the background mesoscale flow \citep{kunze1985}. Ray tracing predicts the NIW frequency to be shifted by $\frac{1}{2}\zeta$. This prediction, however, applies only in regions of the ocean where the WKB limit of ray tracing is appropriate. YBJ argued that this was not the case for the Ocean Storms Experiment, suggesting that the observations were taken in a region where the waves are instead in the so-called strong-dispersion limit. In this limit, refraction of the large-scale wave field is strongly opposed by dispersion, and the frequency shift is much smaller than predicted by WKB theory. This provides a compelling potential explanation of \citeauthor{dasaro1995c}'s (\citeyear{dasaro1995c}) observation that the NIW frequency shift was at least five times smaller than $\frac{1}{2}\zeta$ during Ocean Storms.

This interpretation of the Ocean Storms Experiment was pursued further by \citet{balmforth1998}, who ran spin-down simulations of the YBJ equation. NIWs were initialized in the mixed layer and evolved in the presence of an idealized, barotropic mesoscale eddy field. Qualitative comparisons between the simulations and observations showed that YBJ dynamics were not inconsistent with the observed time for NIWs to escape the mixed layer. \citet{balmforth1999} showed that including the $\beta$-effect improved the agreement with observations. Because an idealized eddy field was used, however, no quantitative conclusions could be drawn about the ability of YBJ to capture the observed evolution.

More recently, \citet{asselin2020penetration} investigated the fate of NIWs as they propagate into a baroclinic mesoscale eddy field using numerical simulations of the extended YBJ system that also accounts for the effect of the waves on the mesoscale. They observed the initially horizontally uniform NIWs undergoing scale reduction by $\zeta$-refraction and then propagate downwards in anticyclones. The vertical wave propagation terminated at depth due to the decay of the baroclinic vorticity away from the surface. For strong NIWs, they also found that the mesoscale eddy field was weakened due to stimulated generation. While this work was motivated by observations, it again employed an idealized setup that made direct comparisons to observations difficult.

In this study, we aim to bridge the gap between theory and observations by using the YBJ framework to interpret the observed evolution of NIWs in the Northeast Atlantic Ocean. We use an array of nine moorings to capture some of the mesoscale variations in the NIW field. The YBJ framework allows us to attribute the observed NIW evolution to a set of well-defined physical processes. We integrate the three-dimensional YBJ equation using observational inputs for the wind forcing, mesoscale streamfunction, and stratification, and we compare these simulations to simpler slab models. We show that the YBJ model offers significant improvements in modeling NIW evolution, without the need for any tuning. We use the YBJ energy budgets to provide a dynamical interpretation of spatial and temporal variations in the NIW field and quantify the relative importance of the various physical processes involved. We also provide an estimate for the importance of stimulated generation in this region.

\section{Observations}\label{obs}

\subsection{NIW Data}

\begin{figure}[t]
 \noindent\includegraphics[scale = 1]{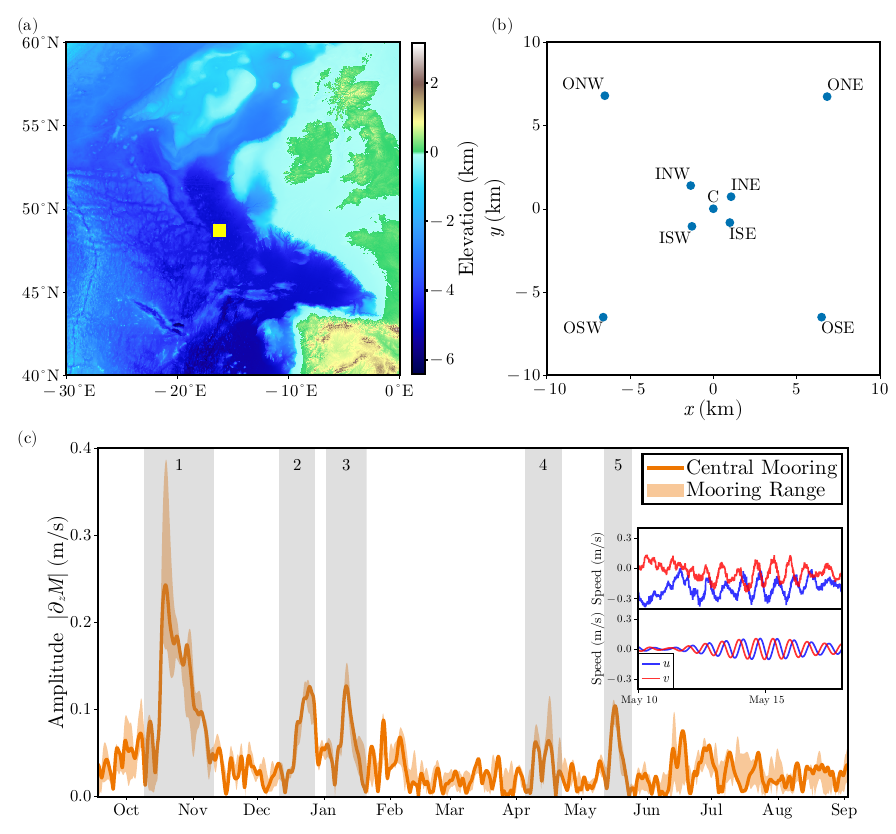}
 \caption{(a)~Location of the OSMOSIS mooring array (yellow square) overlaid on a map of the bathymetry. (b)~Geometry of the mooring array that consists of a central mooring~(C) surrounded by an inner and an outer group of moorings. The inner moorings are labeled with an ``I'' followed by the cardinal direction of the mooring in relation to the central mooring. The outer moorings are labeled in a similar way but with an ``O'' identifier. (c)~Time-series of NIW amplitude extracted from the OSMOSIS mooring observations from September 2012 to September 2013. The solid line indicates the observations at the central mooring while the orange ribbon represents the range across the mooring array. The gray shading indicates the five NIW events discussed in the text. To illustrate the process of extracting the NIW signal, the inset shows the raw velocity (top) and NIW velocity (bottom) during event~5. The amplitude of the envelope modulating the NIW velocity is $|\partial_zM|$.}\label{OSMOSIS}
\end{figure}

We extract observations of NIWs from data collected for the Ocean Surface Mixing, Ocean Submesoscale Interaction Study \citep[OSMOSIS;][]{buckingham2016}. As part of the study, measurements were taken from nine moorings anchored over the Porcupine Abyssal Plain in the northeastern part of the Atlantic Ocean from September 2012 to September 2013. The moorings were all distributed less than 10~km from a central mooring located at ($48.6875^\circ$N, $-16.1875^\circ$E) (Fig.~\ref{OSMOSIS}a,b). In this study, we focus on observations from the central mooring and the four outermost moorings. Each mooring took measurements of the horizontal velocity once every ten minutes using an acoustic current meter (ACM). On the central mooring, there were 13~ACMs spaced nominally between 50~m and 500~m depth. On the outer moorings there were five ACMs nominally spaced over the same depth range. The deepest ACM failed on the outer northeast (ONE) mooring, and so measurements are only available to a depth of 361~m.

The NIW signal is extracted using a Gaussian filter in frequency space. At this latitude, the near-inertial frequency (\qty{0.40}{\per\hour}) is close to the frequency of the M$_2$ tidal constituent (\qty{0.50}{\per\hour}), and so we choose a filter width which corresponds to 11\% of $f$ to exclude the sizable tidal motion from the filtered time series. The conclusions we draw below are not sensitive to the filter width so long as the filter does not include the M$_2$ peak. We identify five events in the year-long time series where strong NIWs were forced relatively coherently across the mooring array (Fig.~\ref{OSMOSIS}c).

The actual depth of the instruments varied in time due to drag exerted by the flow on the moorings. Over the course of the year, there were several times where the moorings experienced knockdown by up to \qty{200}{\meter} \citep{callies2020}. We flag these events if the knockdown on the central mooring is more than \qty{10}{\meter} when averaged with a 1-day running mean. Outside of these knockdown events the variations in mooring depth about the nominal depth is $\sim$\qty{2}{\meter}. The interpretation of the observed NIW signal during these knockdown events is complicated because the filter we use to extract the NIW signal does not commute with evaluating the signal along the trajectory of the ACMs. We do not attempt to explain the observations during the knockdown periods.

\subsection{Stratification and mixed layer Depth}

We need the stratification and mixed layer depth as observational inputs to the numerical YBJ simulations. The stratification influences the dispersion term in the YBJ equation and the mixed layer depth informs us about what depth range to force the waves over. Two ocean gliders sampled across the mooring array during the OSMOSIS study period \citep{damerell2016,thompson2016}. The gliders measured temperature and salinity in the upper 200~m of the water column approximately once every two hours. Following \citet{damerell2020}, we calculate the mixed layer depth~$H_m$ from the glider observations as the depth at which either the potential density~$\rho$ or the temperature~$T$ has changed more than a given threshold from the value at \qty{5}{\meter} depth. These thresholds are $\Delta\rho = \qty{.03}{\kilo\gram\per\meter\cubed}$ for density and $\Delta T = \qty{0.2}{\kelvin}$ for temperature. The final mixed layer depth is taken as the shallowest of the two estimates. 

The stratification~$N^2$ is calculated using:
\begin{equation}
    N^2 = -\frac{g}{\rho}\frac{\partial\rho}{\partial z},
\end{equation}
where $g$ is the acceleration due to gravity. The stratification is then averaged over a given event and the glider trajectories to result in a single spatially and temporally averaged profile used for simulations.

\subsection{Wind Data}

For the wind forcing, we use the European Centre for Medium-Range Weather Forecasting ERA-5 reanalysis \citep{hersbach2018}. We extract time series of the 10~m zonal ($u_w$) and meridional ($v_w$) winds with hourly resolution at the grid point that contained the mooring array. Following \citet{pollard1970} we convert this to a wind stress using a bulk aerodynamic drag formulation. In terms of the complexified wind velocity ($\mathcal{U}_w = u_w + iv_w$), the complexified wind stress~$\tau_w$ is given by:
\begin{equation}
    \tau_w = \rho_aC_D|\mathcal{U}_w|\mathcal{U}_w,
\end{equation}
where $\rho_a$ is the density of air and $C_D$ is the drag coefficient, which we calculate using the speed-dependent formulation of \cite{large1981}.

\subsection{Altimetry}

To characterize the mesoscale eddy field, we use observations of the sea surface height (SSH) from the Data Unification and Altimeter Combination System's (DUACS) delayed-time (DT) 2018 release \citep{taburet2019duacs}. The DUACS DT2018 SSH maps are provided at a nominal $(1/4)^\circ$ and daily resolution. We convert these measurements to a geostrophic streamfunction using $\psi = gh/f$, where $h$ is the SSH and $f$ is the latitude-dependent Coriolis parameter.

\section{Models}\label{models}

\subsection{The PM Model}

We begin by writing the PM model in the language of YBJ. Adding \eqref{eq:PM1} to $i\times \eqref{eq:PM2}$ and multiplying by $e^{-ift}$ yields
\begin{equation}
    \partial_{tz}M = -r\partial_zM + \frac{\tilde{\tau}}{\rho_wH_m},
\end{equation}
where $\tilde{\tau} = e^{-ift}\tau_w$ is the back-rotated, complexified wind-stress. For each NIW event, we solve the PM equation using the reanalysis wind stress and the mixed layer depth from the gliders. In order to better understand the role of $\zeta$-refraction in the life cycle of wind-forced NIWs, we perform a second run of the PM model (denoted by PM+$\zeta$) augmented by the refraction term from the YBJ equation:
\begin{equation}
    \partial_{tz}M = -r\partial_zM -\frac{i\zeta}{2}\partial_zM + \frac{\tilde{\tau}}{\rho_wH_m}.
\end{equation}
This shifts the NIW phase and hence changes the phasing of the NIWs relative to the wind. It does not, however, capture the horizontal energy transfers induced by $\zeta$-refraction because these transfers require horizontal structure in the wave field, which is absent from the horizontally uniform PM model.

For each event, we use a constant mixed layer depth that is an average of the time-varying mixed layer depth over duration of the event. Since the events are relatively short, the error in this approximation is minimal for most events. We make this choice of a constant mixed layer depth to avoid complications that arise otherwise, especially in the YBJ model discussed below.

The parameter~$r$ is intended to account for all of the processes that decrease the wave amplitude in the region of interest. This is primarily thought of as vertical propagation of NIWs out of the mixed layer, but other processes such as advection and dissipation may also cause the NIW amplitude to decrease. Treating all of these processes as a Rayleigh drag term with a single decay parameter represents a drastic simplification in the PM model. It also introduces a free parameter. Previous studies have chosen~$r$ such that the resulting solutions fit observations as closely as possible. The original PM paper used both $r^{-1}=\qty{4}{\days}$ and $r^{-1}=\qty{8}{\days}$. 
\citet{alford2001} used a damping that would correspond to about $r^{-1} = \qty{4}{\days}$ at the latitude of the OSMOSIS mooring array. \citet{yu2022} used $r^{-1}=\qty{16.7}{\days}$ to estimate the NIW wind work during the OSMOSIS experiment using the PM model. We speculate that they had to use very weak damping because their wind data was taken from ERA-interim reanalysis, which has 6-hourly analysis steps with forecasts used to increase the time resolution to 3~hours, which may suppress the wind power at frequencies important for NIW generation. A similarly weak damping is likely unsuitable here, given that ERA-5 reanalysis winds have more power in the near-inertial band \citep[see discussion in][]{flexas2019}. Nevertheless, we vary $r^{-1}$ between \qtylist{4;16}{\days}.

For the PM+$\zeta$ model, we calculate the vorticity using the streamfunction from altimetry. The data processing is as described below for the YBJ model, and we select the vorticity value at the grid point nearest to the center of the mooring array.

We initialize the simulations with no waves and then allow the model to spin up before the main forcing period for each event. We choose the initial time to be when the observations show relatively little waves. This is done by eye. These initial times are followed by a strong forcing event and so the NIW signal is dominated by the newly generated waves, implying that the error from using an initial condition with no waves is relatively small. This discussion also applies to the YBJ simulations we run. We integrate both of the models (PM and PM+$\zeta$) using a Crank--Nicolson scheme.

\subsection{The YBJ Model}

The YBJ equation is a three-dimensional partial differential equation, making it substantially more computationally expensive to integrate than the PM model. We solve the YBJ equation using the pseudospectral solver Dedalus \citep{burns2020} with a mixed explicit and implicit diagonal RK2 scheme. As discussed above, we start the simulations with no waves. We use a domain that is \qtyproduct{400 x 400}{\kilo\meter} in the horizontal (centered on the mooring array) and 4~km deep. Each dimension is discretized with 128~modes. The vertical dimension is finite and represented using Chebyshev polynomials. The horizontal dimensions are made periodic and represented using Fourier modes. The stratification and the wind forcing are taken to be horizontally uniform, capturing the forcing at a scale much larger than the mesoscale. Smaller-scale structure in the wind stress can generate smaller-scale NIWs, but the energy input tends to be strongly dominated by the large-scale winds \citep{rama2022wavelength}. To construct the mesoscale streamfunction used in the simulations, we take $\psi$ from observations and calculate the vorticity by taking the finite-difference Laplacian of $\psi$ on the sphere. We then interpolate this vorticity onto the Cartesian simulation grid centered on the location of the central mooring. We apply a taper to the vorticity, such that it goes to zero on the domain boundaries. We invert this tapered vorticity field for the streamfunction on the periodic domain, making the resulting streamfunction periodic as well. Our conclusions are neither sensitive to the form nor width of the taper. For the results shown below, we use the following form of the taper:
\begin{equation}
    \mathcal{T}(x,y) = \cos^2{\frac{\pi x}{L_x}}\cos^2{\frac{\pi y}{L_y}},
\end{equation}
where $L_x$ and~$L_y$ are the $x$ and~$y$ lengths of the domain, respectively.

The YBJ equation can be formulated on a $\beta$-plane by making the substitution $\zeta/2\rightarrow\zeta/2 +\beta y$. We perform all of our analysis on an $f$-plane (i.e., $\beta=0$) for two reasons: (i)~We focus our analysis on a region in the Northeast Atlantic Ocean where $\beta L \ll \zeta/2$ with $L$ a typical meridional scale of the waves. This relative scaling varies regionally in the ocean. For example, \citet{thomas2020} studied a region that was similarly dominated by the vorticity and found that the $\zeta$-refraction process was more important to the NIW evolution there than $\beta$-refraction, whereas $\beta$-refraction appeared to be important for Ocean Storms \citep{dasaro1995a}. (ii)~The $\beta$ term adds difficulty in simulating the YBJ equation numerically \citep[see][]{balmforth1999}.

For the forcing, we need to prescribe a vertical profile that determines at what depths the momentum is deposited. We specify the profile for the body force $\partial_z\mathcal{F}$ such that it is constant in the forcing layer and then decays rapidly to zero below:
\begin{equation}
    \partial_z\mathcal{F} = \frac{\alpha\tilde{\tau}}{H_m\left[\alpha+\ln(2\cosh\alpha)\right]}\left[1+\tanh{\left(\alpha\frac{z+H_m}{H_m}\right)}\right],\label{eq:forcing}
\end{equation}
where $\alpha$ is a parameter determining the steepness with which the body force falls off below the forcing layer. We use $\alpha = 2$ in all the simulations presented below. The prefactor ensures that $\mathcal{F}=\tilde{\tau}$ at $z=0$.

When the winds blow on the ocean surface, they generate turbulence that mixes momentum downwards. If a mixed layer already exists and is not too deep, the momentum input from the wind will be rapidly homogenized within the mixed layer \citep{pollard1970,kato1969}. For some events at the OSMOSIS site, the mixed layer was up to a few hundred meters deep. These deep mixed layers are likely the result of convection driven by buoyancy forcing rather than the mechanical wind forcing \citep{thompson2016}. In these cases, it is unlikely that the momentum is uniform across the mixed layer, especially if the buoyancy forcing has ceased. The depth structure of NIWs obtained from the OSMOSIS mooring confirms this picture. The waves are initially forced over a layer that is thinner than the mixed layer before they propagate to depth. To avoid forcing over an unrealistically large depth, we cap the forcing layer at \qty{80}{m}. This value is guided by the observations and represents an average depth over which waves are forced when the mixed layer is deep. We discuss possible ways to improve this representation below.

In order to solve the YBJ equation, we must further specify vertical boundary conditions. The requirement that the vertical velocity is zero on the top and bottom of the domain translates to the requirement that $M$ be horizontally uniform at the top and bottom boundaries \citep{young1997}. Since $M$ is determined by $\partial_zM$ up to an arbitrary horizontal function, without loss of generality we specify $M(x,y,-H,t)=0$. The top boundary condition can be found by vertically integrating the YBJ equation twice and requiring that $M$ be horizontally uniform at $z = 0$ (Appendix B). The result is
\begin{equation}
    \partial_tM(x,y,0,t) = \mathcal{F},\label{eq:boundary condition}
\end{equation}
which we integrate a~priori and then use as a Dirichlet boundary condition on~$M$.

\section{Results}

\subsection{Case Study: Fall Event}

\begin{figure}[t]
 \noindent\includegraphics[scale = 1]{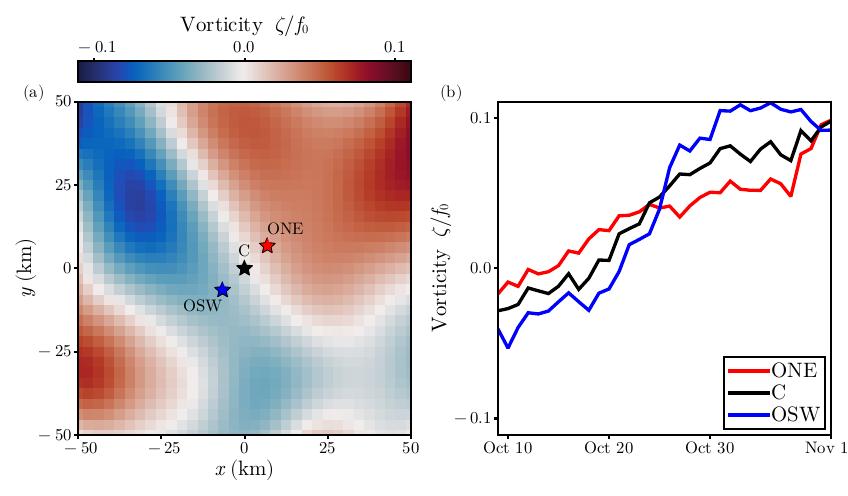}
 \caption{(a)~Snapshot of mesoscale vorticity field in the inner quarter of the simulation domain on 2012-10-18 which was the peak of event~1. The stars show the locations of three specific moorings. The red star denotes the ONE mooring which is in a region of cyclonic vorticity, the black star denotes the central mooring and the blue star denotes the OSW mooring which is in an anti-cyclonic region. (b)~Time series of the vorticity during event~1 at each of the three moorings denoted by the stars above. During the main forcing period the vorticity changes sign across the mooring array while at later times the entire mooring array transitions to being in a region of cyclonic vorticity.}\label{vorticity snapshot}
\end{figure}

We begin with a detailed analysis of the simulation results for event~1, which occurred in the fall (Fig.~\ref{OSMOSIS}c). This event is by far the most energetic NIW event observed throughout the year. The main forcing for event~1 occurred when the mooring array straddled a dipole in the mesoscale vorticity (Fig.~\ref{vorticity snapshot}), making it a good candidate to see the effect of $\zeta$-refraction.

\subsubsection{NIW Amplitude and Phase}

\begin{figure}[t]
 \noindent\includegraphics[scale = 1]{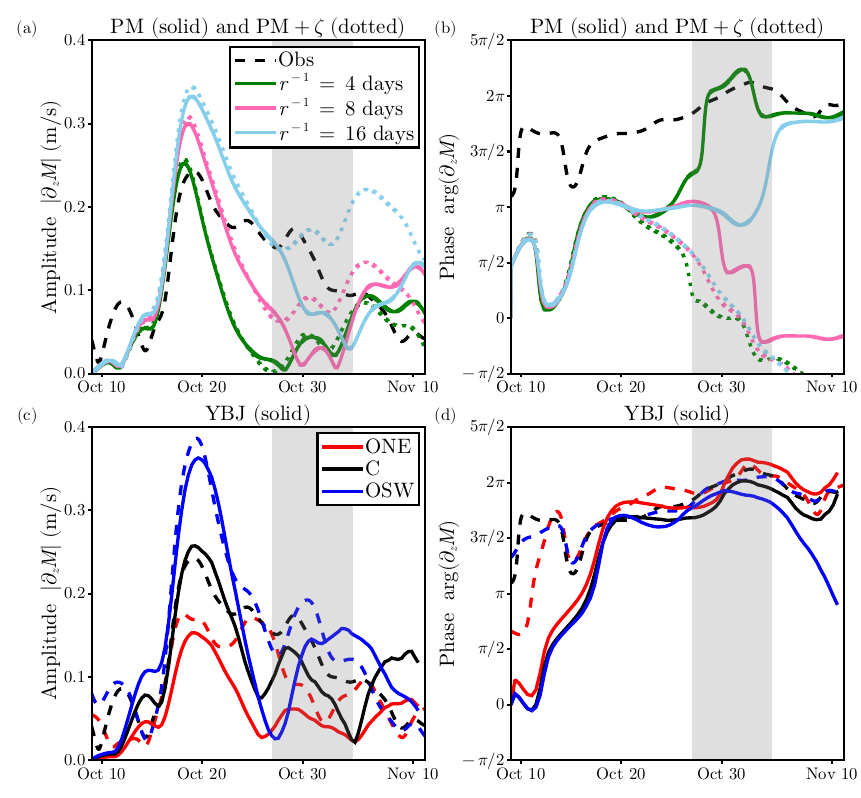}
 \caption{(a) NIW amplitude in the PM model compared to observations at the central mooring (dashed black line). The PM model was run with a range of difference values for the damping parameter $r$. Solid lines indicate the results of the PM model without refraction and dotted lines indicate the results with the refractive term added. The gray shaded region is a period of mooring knockdown. (b) As in (a) but for the NIW phase. (c) NIW amplitude in the YBJ simulation (solid lines) compared to observations (dashed lines) at the central mooring (black) as well as the ONE (red) and OSW (blue) moorings. (d) As in (c) but for the NIW phase.}\label{event 1 results}
\end{figure}

The observed NIW amplitude begins increasing around Oct.~14 and reaches a maximum value on Oct.~20 before returning to background levels by the end of our simulation period on Nov.~10 (Fig.~\ref{event 1 results}a). There are large variations in the peak amplitude over the mooring array, although these differences disappear by the end of the event (Fig.~\ref{event 1 results}c). The phase increases sharply near the beginning of the forcing period as it aligns with the wind (Fig.~\ref{event 1 results}b). Following this, it slowly increases for most of the event and is relatively homogeneous across the mooring array except for a period following the initial forcing (Fig.~\ref{event 1 results}d).

The PM model has trouble capturing the NIW evolution during this event (Fig.~\ref{event 1 results}a,b). Using a value of the damping parameter $r^{-1}=\qty{4}{\days}$ results in a peak amplitude close to that of the observed NIW peak amplitude in the central mooring, but such a relatively strong damping causes the NIW amplitude to drop off much too quickly compared to the observations. A more realistic amplitude decay is achieved when weaker damping is applied (especially with $r^{-1}=\qty{16}{\days}$), but then the peak NIW amplitude is overestimated substantially. The simulated phase bears little resemblance to the observed phase, with the simulated phase being offset by up to $\pi/2$ during the initial forcing period, and the simulated phase remains close to constant around $\pi$ as long as substantial amplitude remains, missing the gradual increase in the observed phase.

The addition of the refractive term does little to change the simulated NIW amplitude in the PM model during most of the event (Fig.~\ref{event 1 results}a,b), indicating that the phase shifts introduced by refraction are unable to substantially change the relative alignment between the NIWs and the winds. At later times the effect of adding refraction is more pronounced in the runs with lower damping as there is more time for refraction to act on the waves before they are dissipated. Refraction seems to dominate the phase evolution at all three damping values, but the phase trends in the opposite direction from what is observed. The phase tendency is also larger in magnitude than what is observed.

While the PM model may be able to capture some qualitative features of the NIW observations, it seems clear that: (i)~the physical processes that cause the NIWs amplitude to decay cannot be accurately captured using a simple linear drag formulation, and (ii)~if refraction is important to NIW evolution, its effect is not simply to change the phase of the NIWs but must involve horizontal processes.

The YBJ model captures the observed amplitude and phase evolution much better than the PM model (Fig.~\ref{event 1 results}c,d). At the central mooring, the YBJ simulation agrees with observations in terms of peak amplitude and decay timescale. Similarly, the phase evolution is much closer to observations than the PM results, in terms of both its value after the forcing and its trend afterward. Again, we emphasize that, unlike the PM models, there is no tunable parameter in the YBJ simulations.

The YBJ model also captures observed lateral variations of the NIW signal across the mooring array (Fig.~\ref{event 1 results}c,d). At the OSW mooring, which at the start of the event is in a region of anti-cyclonic vorticity (Fig.~\ref{vorticity snapshot}), the YBJ simulation successfully predicts a substantial enhancement in the NIW amplitude compared to the central mooring. At the ONE mooring, which at the start of the event is in a cyclonic region (Fig.~\ref{vorticity snapshot}), the YBJ simulation successfully shows a reduction in the NIW amplitude compared to the central mooring. The YBJ model also captures that the NIW phase is much more uniform across the mooring array than the amplitude.

\subsubsection{NIW kinetic energy Budget}

\begin{figure}[t]
 \noindent\includegraphics[scale = 1]{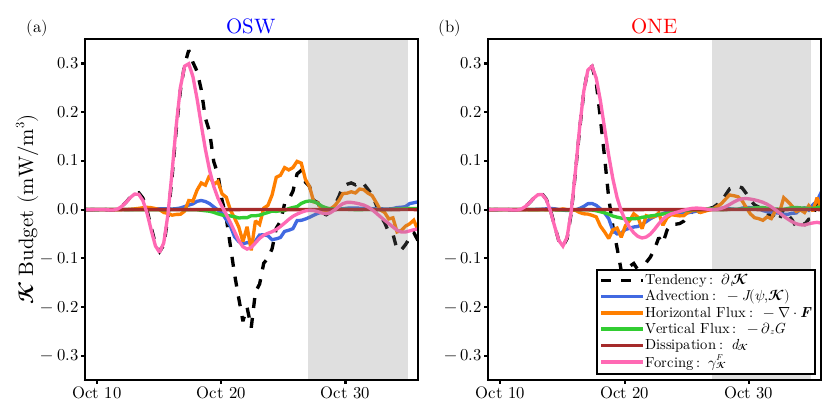}
 \caption{(a)~NIW kinetic energy budget terms. The kinetic energy tendency (dashed line) is decomposed into the 5 processes in the model which can change the kinetic energy: advection (blue), horizontal flux divergence (orange), vertical flux divergence (green), hyperviscosity (brown) and wind forcing (pink). The budgets are evaluated at the horizontal position of the moorings and at fixed depth of \qty{25}{\meter}. (b)~As in (a) but for the ONE mooring. To better visualize the terms we only plot the budget for the first 3/4 of the event.}\label{event1 KE}
\end{figure}

\begin{figure}[p]
 \noindent\includegraphics[scale = 1]{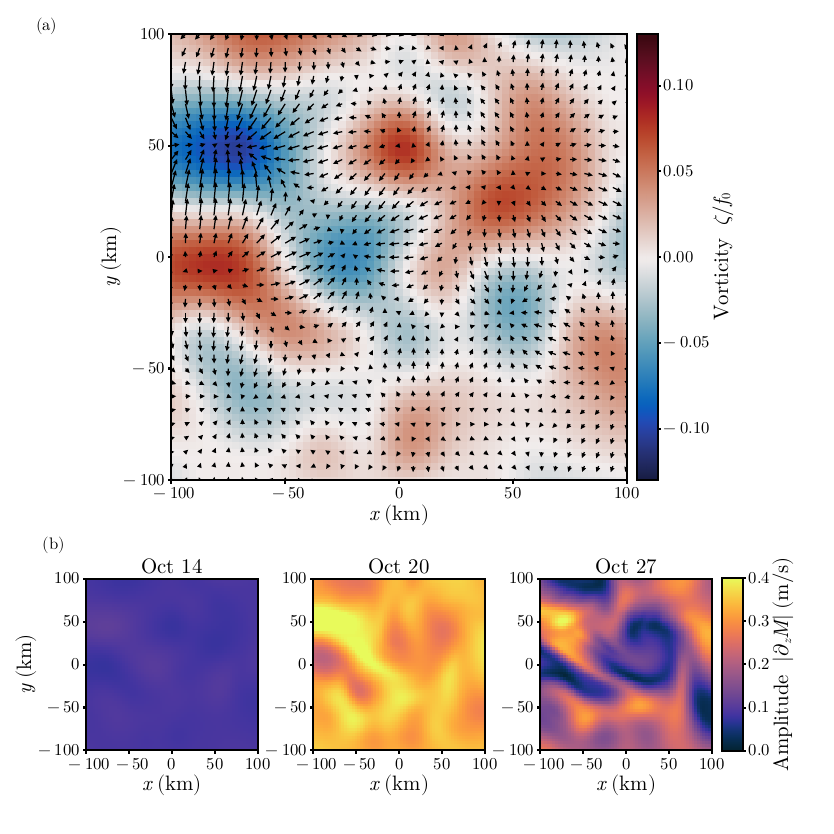}
 \caption{(a)~Color map of vorticity (normalized by $f$) at the peak of event~1. Arrows indicate the horizontal energy flux field at $z=-\qty{25}{\meter}$, showing a flux of NIW kinetic energy out of cyclones and into anticyclones. (b)~Horizontal sections of NIW amplitude at a depth of \qty{25}{\meter} and three different times. Sections show the inner half of the simulation domain. There is no horizontal structure to the forcing and hence the NIWs are initially forced uniformly throughout the domain. Refraction is the only process which can impose structure on a uniform NIW field in the YBJ framework. NIWs begin to be concentrated into anticyclones. Brighter regions in the Oct.~20 plot correlate with anticyclonic regions in the altimetry. Once refraction creates horizontal structure other processes can act. Dispersion will eventually counteract concentration into anticyclones. Advection by the mesoscale eddies will also stir horizontal structure created by refraction. A signal of advective stirring is clearly visible in the upper right quadrant of the Oct.~27 plot.}\label{zeta refraction}
\end{figure}

The observations show a higher NIW amplitude at the mooring in the anticyclone than at the mooring in the cyclone (Fig.~\ref{event 1 results}c,d). This is suggestive of $\zeta$-refraction concentrating NIW kinetic energy in anticyclones. It is hard to draw this conclusion from the observations alone, however, because other factors could give rise to the observed amplitude differences. For example, the current meters were located at slightly different depths (between $z=-\qty{44}{\meter}$ and $z=-\qty{62}{\meter}$).

With the YBJ simulations capturing the observed differences between the moorings, we can use the YBJ framework to identify the processes giving rise to these lateral variations. We construct a point-wise kinetic energy budget~\eqref{eq:KE_budget} at the OSW (anticyclonic region) and ONE (cyclonic region) moorings, which allows us to separate the kinetic energy tendency into advection, dispersive flux divergence, dissipation, and forcing (Fig.~\ref{event1 KE}). During the initial forcing period (up to Oct.~20), the tendency due to wind forcing is similar for both moorings. At both locations, there is a small positive advective tendency that turns slightly negative toward the end of the forcing period. The vertical flux divergence is also similar at the two locations and smaller in magnitude than the advective tendency. Dissipation is negligible. The most notable difference between the two mooring locations is in the horizontal flux divergence term. There is horizontal flux convergence at the OSW mooring and flux divergence at the OSE mooring. This causes the total tendency to be larger than the wind forcing at the OSW mooring and smaller than the wind forcing at the ONE mooring.

At these early times, $\zeta$-refraction is the primary driver of the horizontal energy flux and causes concentration of NIW kinetic energy into anticyclonic regions. During the peak of the event, the horizontal energy flux is directed from cyclonic to anticyclonic regions (Fig.~\ref{zeta refraction}a). This arises from an interplay of the refraction and dispersion terms in the YBJ dynamics: refraction sets up phase gradients, which cause a dispersive energy flux as described by~\eqref{eq:fluxphase}. As horizontal structure develops, advection can also become important as it stirs the existing horizontal structure. This sequence of events was described by \citet{rocha2018} and captures the early evolution during this event (Fig.~\ref{zeta refraction}b).

We again emphasize that the PM model, even if the refractive term is included, cannot capture these dynamics because lateral energy transport originating from the dispersion term in the YBJ equation is crucial. Once dispersion is included, the YBJ model captures observed lateral variations in the NIW amplitude between anticyclonic and cyclonic regions and offers a clear dynamical explanation.

\subsubsection{NIW potential energy Budget}

\begin{figure}[t]
 \noindent\includegraphics[scale = 1]{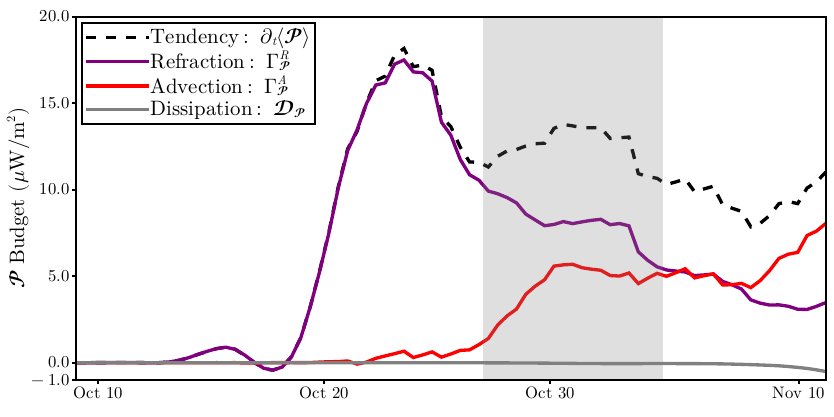}
 \caption{Vertically integrated and horizontally averaged potential energy budget showing the potential energy tendency (dashed line) split into contributions from refractive production (purple), advective production (red) and hyperviscosity (gray).}\label{PE budget}
\end{figure}

The NIW potential energy budget is of interest because it provides insight into the energy exchange with mesoscale eddies \citep{xie2015,rocha2018}. While we prescribe the mesoscale eddy field using altimetry rather than evolving a coupled system, we still interpret the sources of NIW potential energy as estimates of the energy transfer from mesoscale eddies facilitated by NIW refraction and advection.

The vertically integrated and domain-averaged budget for event~1 shows that NIW potential energy is generated by both refraction and advection (Fig.~\ref{PE budget}). The potential energy tendency rises sharply from zero to a peak value after the main forcing period of the event, lagging the peak in the kinetic energy tendency by a few days. At these early times, almost all of the NIW potential energy is created by refractive production. As horizontal structure is created, advective production ramps up. The refractive production term decreases throughout the rest of the event and by the end is overwhelmed by advective production. The initial dominance of refractive production followed by an increase in advective production is very similar to the succession of events \citet{rocha2018} described for stimulated generation in idealized simulations that included the full coupling with the mesoscale dynamics.

The potential energy production---and presumed sink of mesoscale energy---peaks at about \qty{20}{\micro\watt\per\square\meter}. To put this number in context, a global input of \qty{1}{\tera\watt} into the mesoscale eddy field \citep{wunsch2004} corresponds to about \qty{3}{\milli\watt\per\square\meter} on average. For stimulated generation to be important in the global energy budget of mesoscale eddies, it must be much stronger elsewhere.

\subsubsection{NIW Wind Work}

\begin{figure}[t]
 \noindent\includegraphics[scale = 1]{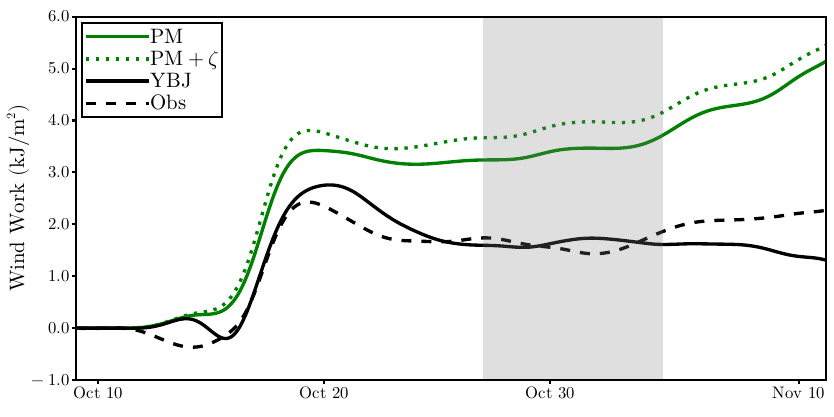}
 \caption{Wind-work as estimated from observations at the central mooring (dashed black), PM (solid green), PM+$\zeta$(dotted green) and YBJ (solid black). For observations the wind-work is defined as $\bf{\tau}_{NIW}\cdot\bf{u}_{NIW}$, while for the simulations it is defined as the kinetic energy production term from the respective kinetic energy budget. The PM models use $r^{-1}=4$~days which best matches the peak amplitude.}\label{wind work}
\end{figure}

The interaction between NIWs and mesoscale eddies also affects the wind work in the near-inertial band (Fig.~\ref{wind work}). The PM and PM+$\zeta$ models with $r^{-1} = \qty{4}{\days}$ best match the peak amplitude in the NIW evolution (Fig.~\ref{event 1 results}a) but overestimate the wind work at the location of the central mooring by a factor of more than two. We could also tune~$r$ to match the wind work estimated directly from observations integrated over the event. This is achieved with $r^{-1} = \qty{0.576}{\days}$, but this means that the peak NIW amplitude is underestimated by a factor of three and violates $r\ll f$. The effect of refraction in the PM model is to slightly increase the wind work. In the YBJ model, in contrast, the wind work matches the observations well. This is a consequence of the YBJ model's ability to closely reproduce the observed NIW evolution. While differences in the wind work between the YBJ and PM models appear substantial, we stress that this event is unlikely to be representative of a time and space average. We intend to discuss this difference further in a subsequent publication.

\subsection{Further Events in the Time Series}

\begin{figure}[p]
 \noindent\includegraphics[scale = 0.93]{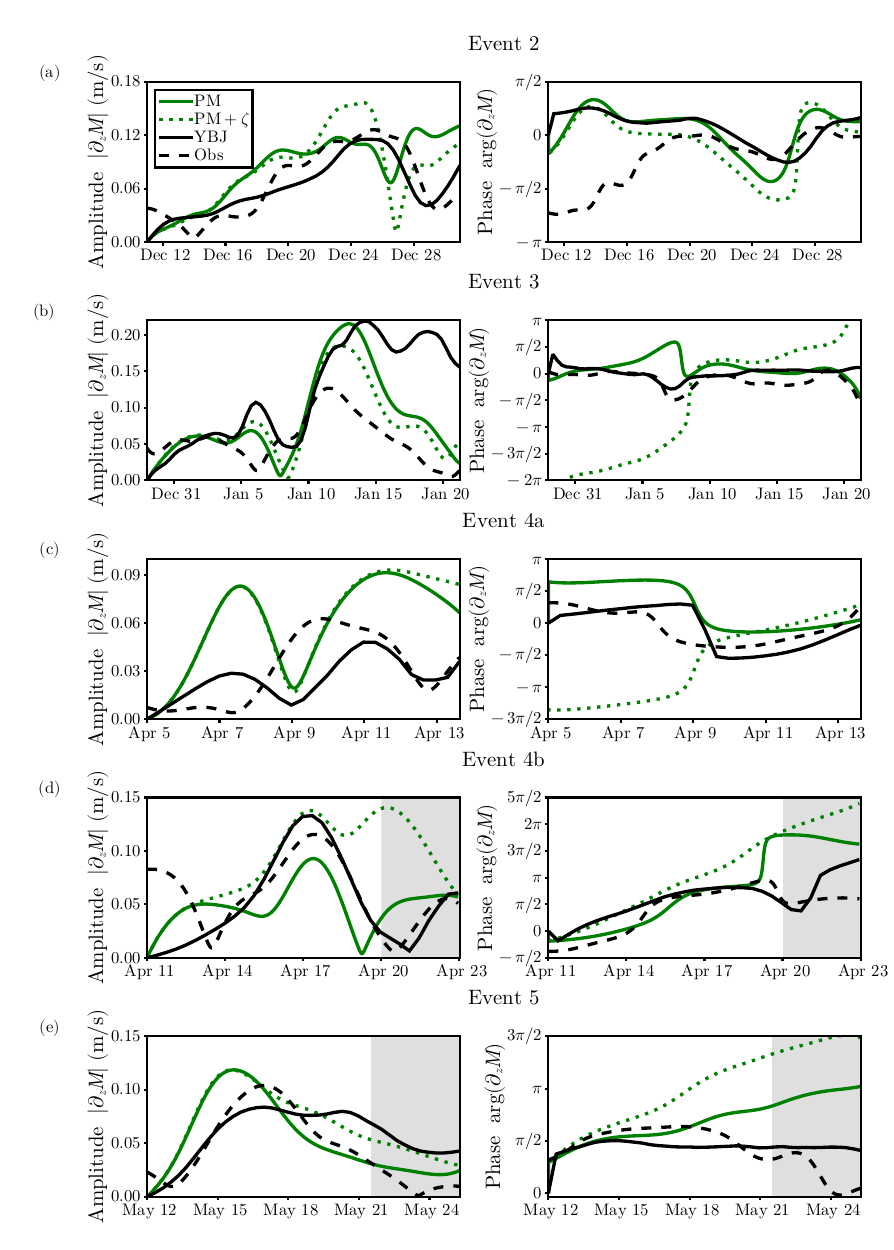}
 \caption{NIW amplitude (left column) and NIW phase (right column) in observations at the central mooring (dashed black line) as compared to PM (solid green) PM+$\zeta$ (dotted green) and YBJ (solid black) for (a)~event~2, (b)~event~3, (c)~event~4a, (d)~event~4b and (e)~event~5.}\label{events2-5}
\end{figure}

Simulations of the remainder of the NIW events (Fig.~\ref{events2-5}) confirm that the YBJ model better captures the observed NIW amplitude and phase evolution than the PM model does (Fig.~\ref{events2-5}). Event~4 is different from the others in that it consists of a double peak in NIW amplitude without a decay to background levels in between. The mixed layer depth was much more variable than during other events, especially towards the end of the event. For that reason, we simulate event~4 in two  parts (events~4a and~4b) in order to minimize the variations in the mixed layer depth over a simulation period.

In general, we see that the YBJ model performs better than the slab models in reproducing the observed evolution of the NIW amplitude and phase. In event~2, the YBJ model captures the slow rise of the NIW amplitude as well as the decrease at late times. The PM models are not able to capture this behavior as well (Fig.~\ref{events2-5}a). All of the models, however, have substantial errors in the phase at early times. This could be due to pre-existing NIWs that we do not capture because we initialize the models at rest. The notable exception to the trend that YBJ performs better than the PM models is event~3 (Fig.~\ref{events2-5}b), where the YBJ model not only overestimates the peak but also the decay time of the waves. YBJ does, however, capture the slow evolution of the phase better than the PM models. We discuss some of the potential reasons for these disagreements below. For event~4a, all models predict a double peak in the amplitude that is not seen in observations (Fig.~\ref{events2-5}c). The YBJ model does better at later times in both amplitude and phase. In event 4b, the YBJ model does rather well in predicting the complete evolution of both the NIW amplitude and phase (Fig.~\ref{events2-5}d). The YBJ model predicts the timing of the peak better than the PM models in event~5, but the waves persist for longer than observed (Fig.~\ref{events2-5}e).


\section{Discussion}

The YBJ model does well in reproducing the mooring observations of NIWs. There are still differences between the model and the observations, however, as well as a couple of events where the YBJ model does less well. Given the observational inputs to the model, it would be surprising if this were not the case. One major limitation is the use of altimetry for the mesoscale streamfunction. As mentioned previously, the altimetry data are a heavily smoothed version of the real field. We suggested above that some of the discrepancies between the YBJ model and the observations were due to this smoothing. We also interpolated the mesoscale vorticity onto a smaller simulation grid. The resolution of the altimetry product is $0.25^\circ\times0.25^\circ$, which is larger than the size of the mooring array. Differences across the mooring array come from the interpolation between neighboring altimetry grid cells, which will have introduced interpolation errors.

Furthermore, we assumed that the mesoscale eddy field was barotropic. This is a reasonable assumption if the vertical scale of the waves is much smaller than the vertical scale of the eddies. There are certainly errors in the YBJ evolution, however, that arise from neglecting the baroclinicity of mesoscale eddies. These effects could be investigated in the future by running similar simulations using an \textit{in-situ} data set that resolves the vertical structure of one or more eddies.

\begin{figure}[t]
 \noindent\includegraphics[scale = 1]{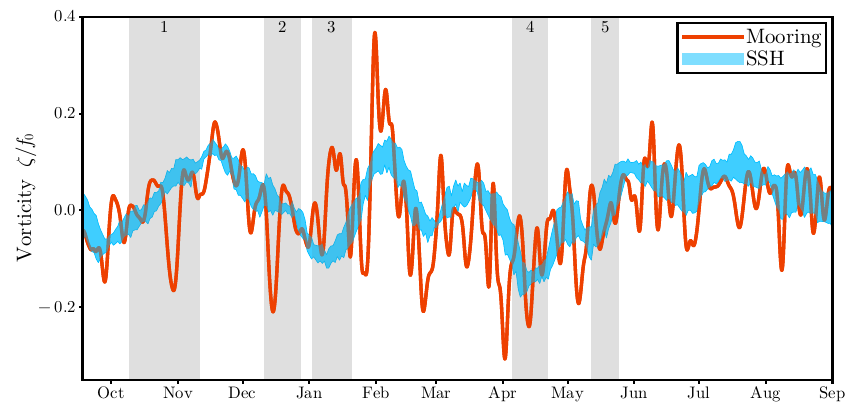}
 \caption{Comparison of the vorticity as calculated from Stokes' theorem applied to the outer moorings (orange) and from altimetry (blue). The velocities used in the Stokes' theorem calculation were obtained by low-pass filtering with a second order Butterworth filter and a cutoff frequency corresponding to a period of approximately 5~days. The blue ribbon shows the spread in the vorticity when interpolated onto the outer mooring positions.}\label{vorticity comparison}
\end{figure}

In event~1, the vorticity has large variations across the mooring array (Fig.~\ref{vorticity snapshot}a). This is reflected in the NIW observations as large differences in the amplitude across the mooring array. However, this need not be the case for all events. Event~3, by contrast, shows weak variations in the vorticity (as diagnosed from altimetry) across the mooring array. The result is that the YBJ simulations also show weak variations in the NIW field across the mooring array region. The altimetry vorticity field is a heavily smoothed representation of the real vorticity field of the ocean; smaller scale vorticity features are invisible to measurements from altimetry. If there is little variation in the larger-scale vorticity across the mooring array, then these smaller-scale vorticity features may play a more important role in governing the variations in~$\partial_zM$. While we do not have a spatial map of smaller-scale vorticity features, we can estimate their magnitude at the mooring array by calculating the vorticity by applying Stokes' theorem to the area bounded by the outer moorings \citep{buckingham2016}. The velocity that we use in Stokes' theorem is low-pass filtered to estimate the balanced signal. This vorticity agrees in general with the vorticity calculated from SSH but shows more high-frequency variability (Fig.~\ref{vorticity comparison}).  Event~3 shows the biggest disagreement between the two estimates of all the events. The mooring estimate shows cyclonic vorticity, whereas the altimetry estimate shows anticyclonic vorticity. This likely explains why we see the amplitude decay much quicker in observations compared to simulations because there would have been a horizontal energy flux out of the region while the simulations have a flux into the region, which acts to maintain the amplitude. The other events show better agreement between the two vorticity estimates, although there are times where the deviation is larger. Events~1 and~2 specifically show two anticyclonic periods that are not captured by altimetry. These may explain some of the mismatch between simulations and observations. For example, the anticyclonic excursion in event~1 probably explains why the YBJ simulation results decay more quickly than observations. For event~2, the excursion occurs primarily near the start of the event, where the NIW amplitude is weak, so its effect is tempered. Many of these anomalies are short in duration, which limits the error in using the altimetric vorticity.

The vorticity error for event~4a is minimal (Figure~\ref{vorticity comparison}). We suspect the mismatch between the YBJ model and the observations in this event is due to the forcing. The reanalysis product used is not the exact wind-forcing felt at the OSMOSIS site. Both YBJ and PM show an initial peak in the NIW amplitude that is not seen in the observations, indicating that the wind-stress may be wrong at this point. The forcing event seen in observations then has to destroy these waves before forcing new ones, which causes the lower amplitude of the NIW peak in the YBJ simulation compared to observations.

Notwithstanding the caveats above, it is significant that the YBJ model can reproduce much of the NIW evolution with only the mesoscale vorticity as derived from altimetry. One may expect the larger-magnitude submesoscale vorticity to be at least as important for the NIW evolution. But the dispersion term in the YBJ equation depends on the Laplacian of~$M$ and hence in spectral space scales as $\kappa^2$. This means that the refractive generation of small-scale structure in the wave field will be opposed by increasingly strong dispersion. Our results therefore suggest that, at the location of the mooring array, dispersion indeed outpaces refraction at submesoscales and mesoscale refraction is more important for the NIW evolution \citep[cf.,][]{yu2022}.

Using a time-averaged stratification profile likely also contributes to differences between the observations and simulations, despite our attempts to keep simulation times as short as possible to avoid this effect. The assumption that the stratification does not  vary in time is inherent in the YBJ scaling assumptions. Related to this is the question of what depth to force the model over. We used a fixed forcing depth, although this certainly varied over the simulation periods. This problem could be side-stepped by representing the wind forcing as a surface stress and adding a parameterized turbulent vertical momentum flux to the YBJ equation to transfer momentum from the wind downwards. 

The YBJ equation contains no term to represent the breaking of NIWs. The scaling $H^*=u^*/\sqrt{N^*f}$, where $u^*=\sqrt{\tau/\rho_0}$ is the friction velocity and $N^*$ is the stratification at the base of the mixed layer, indicates the depth to which wind-driven turbulence could deepen the mixed layer \citep{pollard1973}. If the mixed layer depth is shallower than~$H^*$, the effect of NIW dissipation would be important as there is potential for mixed layer deepening by breaking of NIWs \citep{plueddemann2006observations}. During the OSMOSIS study, \citet{yu2022} showed that the mixed layer was generally deeper than~$H^*$. This is especially true outside of summer and during the events we consider here. This is due to strong surface buoyancy forcing deepening the mixed layer by convection to a depth beyond what the winds alone could do. While one could include a critical Richardson number criterion to add NIW breaking to the YBJ equation, we believe that it is justifiable to ignore dissipation of NIWs in the surface layer in this study.

In event~1, the wind work predicted by YBJ agreed well with that calculated by observations. The PM model greatly overestimated the wind-work. The YBJ and PM models also disagree on the wind work for all the other events (not shown). If mesoscale interactions change the alignment of the NIWs with the winds, then this results in a very different wind work. The difference in wind work between the YBJ and PM models depends on the degree to which the mesoscale changes the alignment of the waves with the wind and the power of the winds at different frequencies. The differences in wind work can be large for individual events but the five events we simulated are not enough to estimate what the average difference might be when averaged over many events. Nonetheless, this may provide motivation to revisit previous estimates of the NIW wind work in order to determine whether the mesoscale may modulate wind work into the NIW band globally.


\section{Conclusion}

The evolution of the NIW field at the OSMOSIS site in the Northeast Atlantic Ocean is strongly modulated by mesoscale eddies. The observed evolution can be captured by the YBJ model, which includes NIW refraction, advection, and dispersion. If these processes are omitted as in the PM model, the observations cannot be reproduced as well, even if the parameterized damping rate is tuned.

The YBJ model provides a powerful interpretive framework for understanding observations of NIWs in the upper ocean. It allows us to attribute the observed evolution to specific physical processes. Lateral differences in the NIW amplitude across the mooring array are caused by $\zeta$-refraction, which causes NIW kinetic energy to be fluxed into anticyclonic regions. While correlations between NIW amplitudes and mesoscale vorticity can be diagnosed from the observations alone \citep[e.g.,][]{yu2022}, the YBJ framework produces a quantitative prediction for the NIW amplitudes that matches observations and provides a physical interpretation.

The YBJ model also allows us to calculate changes in the NIW potential energy, which are expected to arise from transfers of balanced mesoscale energy \citep{xie2015,rocha2018}. For the strongest NIW event observed during the OSMOSIS campaign, the NIW potential energy gain is at least two orders of magnitude smaller than the global average energy input into mesoscale kinetic energy. Stimulated generation therefore does not appear to have a major impact on the mesoscale eddy field in this part of the ocean.

\acknowledgments
The authors thank two anonymous reviewers whose comments helped improve the presentation of this work. This material is based upon work supported by the National Science Foundation under Grant No.~OCE-1924354 and by the National Aeronautics and Space Administration under Grant No.~80NSSC22K1445 issued through the Science Mission Directorate (Future Investigators in NASA Earth and Space Science and Technology).

%
%
\datastatement
The code to run the 3D YBJ model is available at \url{https://github.com/scott-conn/3DYBJ}. All of the OSMOSIS data used as part of this study is available from the British Oceanographic Data Centre. The OSMOSIS mooring data is available at \url{https://www.bodc.ac.uk/data/bodc_database/nodb/data_collection/6093/}. The OSMOSIS glider data is available at \url{https://doi.org/10.5285/6cf0b33e-a192-549f-e053-6c86abc01204}. The SSH data is available from the E.U.'s Copernicus Marine Service at 
\url{https://doi.org/10.48670/moi-00148}. The ERA5 reanalysis data is available from the Copernicus Climate Change Service (C3S) Climate Data Store at \url{https://doi.org/0.24381/cds.adbb2d47}.

\clearpage

\appendix[A]\label{A}
\appendixtitle{YBJ kinetic energy Budget}

\noindent The YBJ kinetic energy budget may be formed by multiplying \eqref{eq:standard_YBJ} by $-M^*/2$ and adding the complex conjugate. Using integration by parts, the resulting kinetic energy budget can be written as
\begin{equation}
    \partial_t\mathcal{K} + J(\psi,\mathcal{K}) + \nabla\cdot\bm{F}'_\mathcal{K} + \partial_zG'_\mathcal{K} = d_\mathcal{K} + \gamma_\mathcal{K}^F,\label{eq:KE_budget_notgaugefixed}
\end{equation}
where
\begin{subequations}
    \begin{align}
        \mathcal{K} &= \frac{1}{2}|\partial_zM|^2,\\
        \bm{F}'_\mathcal{K} &= \frac{iN^2}{4f}\left(M\nabla M^* - M^*\nabla M\right),\label{eq:F}\\
        G'_\mathcal{K} &= -\frac{M^*}{2f}\left(\partial_{zt}M+J(\psi,\partial_zM)+\frac{i\zeta}{2}\partial_zM+\nu \nabla^4\partial_zM-\partial_z\mathcal{F}\right)+\text{ c.c. },\\
        d_\mathcal{K} &= \frac{\nu}{2}\left(\partial_zM^*\nabla^4 \partial_zM + \partial_zM\nabla^4\partial_zM^*\right),\\
        \gamma_\mathcal{K}^F &= \frac{1}{2}\left(\partial_zM^*\partial_z\mathcal{F}+\partial_zM\partial_z\mathcal{F}^*\right).
    \end{align}
\end{subequations}
In this form, however, the vertical flux does not vanish on the boundaries at $z = 0$ and $z = -H$. In order to meaningfully separate horizontal and vertical fluxes, we redefine the 3D kinetic energy flux vector $\mathbf{H}' = [\mathbf{F}',G']^\mathrm{T}$, taking advantage of the fact that only its divergence appears in~\eqref{eq:KE_budget_notgaugefixed}. The kinetic energy budget is thus invariant under the transformation $\mathbf{H}'\rightarrow\mathbf{H} = \mathbf{H}'+\nabla_3\times\bm{\chi}$ with some vector field~$\bm{\chi}$. We propose to pick a~$\bm{\chi}$ such that the transformed~$G$ is zero on both boundaries. To determine~$\bm{\chi}$, we first integrate \eqref{eq:standard_YBJ} from~$-H$ to~$z$ and then introduce a new field~$A$ such that $\partial_zA = N^2M/f^2$, which allows us to evaluate the integral of the dispersion term. The resulting equation is
\begin{equation}
    \partial_{zt}M+J(\psi,\partial_zM)+\frac{if}{2}\nabla^2A+\frac{i\zeta}{2}\partial_zM - \partial_z\mathcal{F}+\nu\nabla^4\partial_zM = \mathcal{C}
    \label{eq:YBJ_vertically_integrated}
\end{equation}
with
\begin{equation}
    \mathcal{C}(x,y,t) = \partial_{zt}M+J(\psi,\partial_zM)+\frac{if}{2}\nabla^2A+\frac{i\zeta}{2}\partial_zM - \partial_z\mathcal{F}+\nu\nabla^4\partial_zM \quad \text{at} \quad z = -H.
    \label{eq:C}
\end{equation}
We are free to choose~$\mathcal{C}$ as a boundary condition on the new field~$A$, i.e., we can choose $A$ at $z = -H$ as the solution of~\eqref{eq:C} for some specified~$\mathcal{C}$. A sensible choice for the transformation is thus setting $\mathcal{C}=0$ and
\begin{equation}
    \bm{\chi} = \frac{if}{4}\begin{pmatrix}
    M^*\partial_yA-M\partial_yA^*\\
    -M^*\partial_xA+M\partial_xA^*\\
    0
    \end{pmatrix}.\label{eq:chi}
\end{equation}
Under this transformation, the kinetic energy budget becomes \eqref{eq:KE_budget}, with the horizontal and vertical fluxes given by
\begin{subequations}
    \begin{align}
        \mathbf{F}_\mathcal{K} &= \frac{if}{4}\left(\partial_zM^* \nabla A - \partial_zM \nabla A^*\right),\\
        G_\mathcal{K} &= \frac{if}{4}\left(\nabla M\cdot\nabla A^* - \nabla M^*\cdot\nabla A\right).
    \end{align}
\end{subequations}
It is clear that the vertical flux~$G_\mathcal{K}$ vanishes at the boundaries, as desired, because $\nabla M = 0$ there.

The divergence of the flux terms vanishes when the budget is integrated over the entire domain, and the only terms that can alter the domain-integrated kinetic energy are the wind work and dissipation:
\begin{equation}
    \partial_t\langle\mathcal{K}\rangle = \mathcal{D}_\mathcal{K}+\Gamma_\mathcal{K}^F,\label{eq:domain_int_KE_budget}
\end{equation}
where $\mathcal{D}_\mathcal{K} = \langle d_\mathcal{K}\rangle$ and $\Gamma_\mathcal{K}^F = \langle \gamma_\mathcal{K}^F\rangle$.

\clearpage

\appendix[B]
\appendixtitle{YBJ Upper Boundary Condition}

\noindent Beginning from~\eqref{eq:YBJ_vertically_integrated} with the choice $\mathcal{C}=0$, we can vertically integrate from $z=-H$ to $z=0$ and use $M=0$ at $z = -H$ to arrive at
\begin{equation}
    \left[\partial_tM + J(\psi,M) + \frac{i\zeta}{2}M - \mathcal{F} + \nu\nabla^4M\right]_{z=0} = -\frac{if}{2}\nabla^2\int_{-H}^0 A \, dz.
    \label{eq:bc1}
\end{equation}
The no-normal flow condition is imposed by requiring~$\nabla M = 0$ at $z = 0$ \citep{young1997}, which eliminates the advection and dissipation terms. We then horizontally average (denoted by $\overline{\,\cdot\,\vphantom{a}}$) equation~\eqref{eq:bc1}. Because $M(x,y,0,t)$ has no horizontal structure, it is equal to its horizontal average. On a horizontally periodic domain, all but two terms vanish in the averaged equation:
\begin{equation}
    \partial_tM(x,y,0,t) = \overline{\mathcal{F}}(0,t).
    \label{eq:bc2}
\end{equation}
Because the forcing is horizontally uniform in all of our simulations, this reduces to~\eqref{eq:boundary condition}. Note that subtracting~\eqref{eq:bc2} from~\eqref{eq:bc1} yields a condition on the integral of~$A$:
\begin{equation}
    \frac{if}{2}\nabla^2\int_{-H}^0A\,dz= \mathcal{F}' -\frac{i\zeta}{2} M(x,y,0,t),\label{eq:B3}
\end{equation}
where $\mathcal{F}'=\mathcal{F}-\overline{\mathcal{F}}$. Note that unlike in YBJ where the integral in left-hand side of Equation~\ref{eq:B3} is set to zero which eliminates the barotropic mode, our boundary conditions allows for a barotropic mode. 

\clearpage

\appendix[C]
\appendixtitle{YBJ Potential Energy Budget}

\noindent The YBJ potential energy budget can be formed by multiplying \eqref{eq:standard_YBJ} by $i\partial_tM^*/2f$ and adding the complex conjugate. However, a more transparent derivation begins with dotting the gradient of~\eqref{eq:YBJ_vertically_integrated} with~$-\nabla A^*/4$ and adding the complex conjugate (setting $\mathcal{C} = 0$ as before). The resulting potential energy budget is:
\begin{equation}
    \partial_t\mathcal{P} + J(\psi,\mathcal{P}) + \nabla\cdot\mathbf{F}_\mathcal{P} + \partial_zG_\mathcal{P} = d_\mathcal{P}+\gamma_\mathcal{P}^F + \gamma_\mathcal{P}^R + \gamma_\mathcal{P}^A,
    \label{eq:PE_budget}
\end{equation}
where
\begin{subequations}
    \begin{align}
        \mathcal{P} &= \frac{N^2}{4f^2}|\nabla M|^2,\\
        \mathbf{F}_\mathcal{P} &=\frac{if}{8}\left[(\nabla^2A^*)\nabla A - (\nabla^2A) \nabla A^*\right],\\
        G_\mathcal{P} &=-\frac{1}{4}\left[\nabla A^*\cdot\left(\partial_t\nabla M + J(\psi,\nabla M)\right) + \text{c.c.}\right],\\
        d_\mathcal{P} &=\frac{\nu}{4}\left[\nabla A^*\cdot \nabla^4 \nabla \partial_z M + \nabla A\cdot \nabla^4 \nabla \partial_z M^*\right],\\
        \gamma_\mathcal{P}^F &=-\frac{1}{4}\left[\nabla A^*\cdot\nabla \partial_z\mathcal{F} + \nabla A\cdot\nabla \partial_z\mathcal{F}^*\right],\\
        \gamma_\mathcal{P}^R &=\frac{i}{8}\left[\nabla A^*\cdot\nabla (\zeta\partial_zM) - \nabla A\cdot\nabla (\zeta\partial_zM^*)\right],\\
        \gamma_\mathcal{P}^A &=\frac{1}{4}\left[\nabla A^*\cdot J(\nabla \psi,\partial_zM) + \nabla A\cdot J(\nabla \psi,\partial_zM^*)\right].
    \end{align}
\end{subequations}
By making a plane wave ansatz in the vertical and using integration by parts, this energy budget can be brought into the same form as the potential energy budget in \cite{rocha2018}. In this paper we are primarily concerned with the domain-integrated potential energy budget. Under domain integration terms which differ by a divergence are the same and hence the interpretation of terms also remains the same. The domain-integrated potential energy budget is~\eqref{eq:domain_integrated_PE_budget}, where again $\mathcal{D}_\mathcal{P} = \langle d_\mathcal{P}\rangle$ and $\Gamma_\mathcal{P}^i = \langle\gamma_\mathcal{P}^i\rangle$. In the case of horizontally uniform forcing, there is no generation of potential energy by the winds and hence $\gamma_\mathcal{P}^F = 0$.

\clearpage





%
%
%
\bibliographystyle{ametsocV6}
\bibliography{references}

%

%

\end{document}